\documentclass[twocolumn,showpacs,preprintnumbers, amsmath,amssymb]{revtex4}
\usepackage{graphicx}
\usepackage{bm}

\begin{document}
\title{Reading-out the state of a flux qubit by Josephson transmission line solitons}

\author{Arkady Fedorov}
\email{arkady@tfp.uni-karlsruhe.de}
\author{Alexander Shnirman}
\author{Gerd Sch\"on}
\affiliation{Institut f\"ur Theoretische Festk\"orperphysik
and DFG-Center for Functional Nanostructures (CFN),
Universit\"at Karlsruhe, D--76128 Karlsruhe, Germany}

\author{Anna Kidiyarova-Shevchenko}
\affiliation{Microtechnology and Nanoscience Department, Chalmers
University of Technology, 412 96 Gothenburg, Sweden}

\begin{abstract}
We describe the read-out process of the state of a Josephson flux qubit via
solitons in Josephson transmission lines (JTL) as they are in use in the standard rapid single flux quantum (RSFQ) technology. 
We consider the situation where the information
about the state of the qubit is stored in the time delay of the soliton. We analyze dissipative underdamped JTLs, 
take into account their jitter, and provide estimates of the measuring time and efficiency of the measurement for relevant experimental
parameters.
\end{abstract}
\pacs{03.67.Lx, 85.25.-j}

\maketitle
\section{Introduction}

The last few years have brought substantial breakthroughs in experiments
with Josephson qubits, with further progress depending to a large extent on
our ability to control the circuits with high precision and, at the same time, avoid decoherence.
One of the promising ideas combines Josephson flux
qubits with the well developed classical rapid
single flux quantum (RSFQ) technology~\cite{Averin,Kaplunenko}. RSFQ elements should allow for more reliable on-chip control and measurements than what can be achieved with control pulses sent over long coaxial cables.
However, they have the drawback to be dissipative and thus noisy. The effect of this noise needs to be investigated, and ways need to be found to minimize it.

\begin{figure}
\includegraphics[width=7cm]{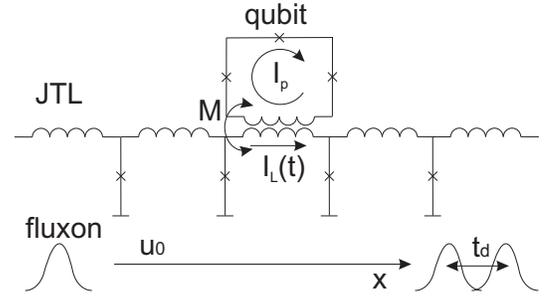} 
\caption{Setup for the read-out of the persistent
current qubit based on the delay time of a soliton in the
Josephson transmission line (JTL).}
\label{fig1}
\end{figure}

In this article we consider one of the possible elements of an RSFQ circuit,
namely a Josephson transmission line (JTL). The JTL supports propagating
signals in the form of Josephson solitons (phase slips),
also called fluxons. As suggested by Averin {\sl et al.}~\cite{Averin} ballistic fluxons in the JTL can be used to read out the state of a superconducting flux qubit in a setup as shown in Fig.~\ref{fig1}. 
In the proposed
schemes the information about the state of a qubit is contained either in the
fluxon transmission probability (transmission detection mode) or
propagation time (delay time detection mode). Under certain ideal conditions the measurement time
is equal to the back-action dephasing time, which proves that the JTL can operate as an ideal detector.
In this paper we 
investigate the efficiency that can be achieved in the delay time detection mode.
For that purpose we evaluate the delay time
in three different setups: a) when the
qubit is kept away from the symmetry point all the time; b) when the qubit is
initially prepared at the symmetry point, but the approaching soliton pushes the qubit far from the symmetry point;
c) when the qubit is near the symmetry point all the
time.  We analyze the relation between the delay time and the
dissipation as well as the probability of errors introduced by the
measurement. Finally, we compare the delay time with the
characteristic time uncertainty due to jitter (thermal fluctuations) in the JTL, and we determine how many solitons are needed for a reliable measurement.

For a flux qubit operated far from the symmetry point the eigenstates are persistent current states. The persistent current in the qubit loop induces an external magnetic flux in the JTL, which provides a scattering potential for the fluxon and is responsible for the time delay of the fluxon propagation. The sign of the magnetic flux and the value of the delay time depend on the state of the qubit, which allows measuring its quantum state.

For a qubit prepared in one of the energy eigenstates at the symmetry point the expectation value of the current in the loop vanishes. However, for strong qubit-JTL coupling the fluxon shifts the working point of the qubit away from the symmetry point, and then again the qubit produces a magnetic flux in the JTL. The delay time is approximately the same as if the qubit would have been in a persistent current state corresponding to the induced shift.

For weak qubit-JTL coupling a fluxon shifts the working point of the qubit only slightly around the symmetry point. For this case we demonstrate that the qubit introduces an effective inductance to the JTL with sign depending on the qubit eigenstate. The local change of the inductance of the JTL also serves as a scattering potential for the fluxons, leading to a delay in the fluxon propagation time. This property can in principal be used for a measurement of the qubit at the symmetry point even for weak qubit-JTL coupling.

\begin{figure}[t]
\includegraphics[width=7cm,trim=1cm 0.5cm 1.0cm 0.5cm]{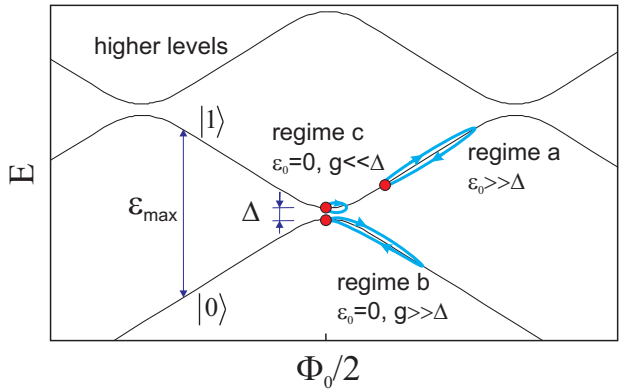}
\caption{Energy diagram of the qubit for the considered measurement schemes: a) The qubit is kept away from the symmetry point. b) The qubit is
initially prepared at the symmetry point, but the approaching soliton pushes the qubit far from the symmetry point.  c) The qubit remains near the symmetry point all the time. } \label{regimes}
\end{figure} 

For a realistic assessment of the feasibility for the proposed measurement schemes we need to consider the major sources of errors. One of them 
is thermal noise in the JTL, as a result of which fluxons show fluctuations in their velocity, which lead to an uncertainty (jitter) in the propagation time. In order to distinguish different eigenstates of the qubit the induced delay time should exceed this time jitter.
Another source of errors are non-adiabatic transitions of the qubit caused by the moving fluxons.
This back-action effect of the JTL on the qubit needs to be taken into account if the qubit is measured at the symmetry point.
Still another source of errors are intrinsic relaxation processes of the qubit due to noise not related to the JTL. The relaxation destroys the state of the qubit to be measured and should be much slower than the measurement time.
Finally, in order to be detectable the delay times should exceed the time resolution of the RSFQ detector used for delay time measurement.

Our calculations show that for suitable JTL parameters the induced delay times well exceed the time resolution of the RSFQ detector.
For strong qubit-JTL coupling the qubit measurement by a single fluxon at the symmetry point, as well as far from it, can be performed with an
accuracy of $70-90\%$ for experimentally accessible parameters.
For high velocities of the fluxon the measurement efficiency is mostly
limited by the jitter and, if the qubit is at the symmetry point, by Landau-Zener transitions.
For low velocities the intrinsic relaxation of the qubit may be important.
To improve the accuracy one has to use many fluxons for the measurement.
If one optimizes the fluxon velocity and number the qubit can be measured with accuracy reaching $99\%$
far from the symmetry point and above $90\%$ at the symmetry point for strong qubit-JTL coupling. The measurement time remains approximately 4~ns for both cases. For weak qubit-JTL couping the qubit prepared at the symmetry point stays there during the whole measurement. In this case the measurement of the qubit is not feasible for standard JTL parameters.

\section{The model}

We focus on the readout of a persistent current
qubit~\cite{PCQ1,PCQ2} via its coupling to solitons in a Josephson
transmission line (JTL) in a setup as shown in
Fig.~\ref{fig1}. The qubit consists of a  superconducting loop with three
Josephson junctions and is inductively coupled to the JTL. The
Hamiltonian of the total system is
\begin{equation}\label{Htotal}
H=H_{qb}+H_{JTL}+H_I,
\end{equation}
where $H_{qb}$ describes the qubit, $H_{JTL}$ the JTL, and $H_I$ the qubit-JTL coupling.

In the basis of the two lowest persistent current states, $|0\rangle$, $|1\rangle$, corresponding to  currents $\pm I_p$ circulating in the qubit loop in opposite directions, the qubit Hamiltonian can be expressed in terms of Pauli matrices~\cite{PCQ2},
\begin{equation}\label{total H1}
H_{qb}=-\frac{\epsilon_0}{2}\sigma_z-\frac{\Delta}{2} \sigma_x.
\end{equation}
The tunneling amplitude $\Delta$, leading to transitions between both states, is fixed by the experimental setup. The energy bias $\epsilon_0$ between the two persistent current states depends on the deviation of the external
magnetic flux $\Phi$ in the loop from the symmetry point,
\begin{equation}\label{epsilon}
\epsilon_0=2I_p(\Phi-\Phi_0/2),
\end{equation}
where $\Phi_0$ is the magnetic flux quantum. 

\begin{figure} 
\includegraphics[width=7cm]
{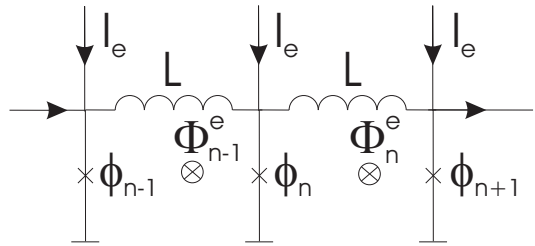} \caption{Two elementary cells of the discrete JTL.} \label{cells}
\end{figure}

The Hamiltonian of the discrete dissipationless JTL, shown in Fig.~\ref{cells}, in vanishing external magnetic field can be expressed in terms of the charges $q_n$ and the phase differences $\phi_n$ of the $n^{\rm th}$ Josephson junctions
\begin{eqnarray}
H_{JTL}&=&\sum_{n=1}^N \left(\frac{q_n^2}{2C}+E_J (1-\cos \phi_n)\right.\nonumber\\
& &+\left.\left(\frac{\Phi_0}{2\pi}\right)^2\frac{(\phi_{n+1}-\phi_n)^2}{2L}-\frac{\hbar I_e}{2 e}\phi\right).
\label{H_JTL}
\end{eqnarray}
Here,  $N$ is the total number of junctions in the JTL, which here are assumed to be equal, $C$ is the capacitance, and $E_J=\hbar I_c/2e$ the Josephson energy of the junction with $I_c$ being the junction critical current.
The inductance of each cell of the JTL is denoted by
$L$, and $I_e$ is the bias current supplied externally to each junction.

The dimensions of the qubit are much smaller than the length  $a$ of a unit cell of the JTL. Therefore, one can assume that the qubit is inductively connected only to  one cell of the JTL denoted by label $m$. The corresponding mutual inductance is $M=k\sqrt{L_{qb}L}$, where $L_{qb}$ is the inductance of the qubit loop and $k$ is the coefficient of coupling.
Thus the Hamiltonian of the qubit-JTL interaction  in the linear response regime is given by
\begin{equation}\label{H_I}
H_I=\sigma_z M I_p I_L =-\frac{g}{2}\sigma_z(\phi_m-\phi_{m+1}),
\end{equation}
where $I_L$ is the electrical current in  the inductance of the $m$-th JTL  cell and $g=2I_p M\Phi_0/(2\pi L)$ is the effective qubit-JTL coupling strength.

The classical dynamics of the uniform, discrete, dissipative JTL is governed by Kirchhoff circuit equations~\cite{Likharev}
which lead to a system of $N$ discrete sine-Gordon equations
\begin{eqnarray}\label{discrete sine-Gordon}
\frac{\ddot \phi_n}{\omega_p^2} +\frac{\dot \phi_n}{\omega_c}+\sin\phi_n
&=&\frac{I_e}{I_c} +\frac{1}{L}(\Phi^e_{n-1}-\Phi^e_n)\\\nonumber
& &-\frac{\Phi_0}{2\pi L I_c}\left(\phi_{n-1}-2\phi_n+\phi_{n+1}\right).
\end{eqnarray}
Here, $\omega_p=(2 e I_c/\hbar C)^{1/2}$ is the plasma frequency of the junction,
$ \omega_c=2 e I_c/\hbar G_N$ the junction characteristic frequency with
$G_N$ being the normal conductance of the junction determining the dissipation in the system, and $\Phi^e_n$ is the external magnetic flux in the $n-$th cell induced by the qubit.

For low inductances, $L I_c\ll\Phi_0$,  and weak external magnetic fields, $\Phi_n^e\ll\Phi_0$, the phases $\phi_n$ of the neighboring junctions are close to each other, and we can rewrite (\ref{discrete sine-Gordon}) in dimensionless differential form,
\begin{equation}\label{sine-Gordon perturbed}
\ddot \phi -\phi_{xx} +\sin \phi=j_e-\alpha \dot \phi+f^{qb}(\phi,x).
\end{equation}
Here the time $t$ and continuous space variable $x$ are measured in the units of the inverse plasma
frequency $\omega_p^{-1}$ and the Josephson penetration depth
$\lambda_J=a\left(\Phi_0/(2\pi L I_c)\right)^{1/2}$, respectively, and the phase difference $\phi(x)$ is a function of $x$.  Furthermore,  $\alpha=\omega_p/\omega_c$ is the
dissipation strength due to a non-vanishing normal conductance $G_N$, and $f^{qb}(\phi,x)$ is the perturbation induced by the qubit.

In the following sections we will
show that far from the symmetry point, $\epsilon_0\gg\Delta$,  the perturbation is created by the magnetic flux due to the qubit and can be expressed as
\begin{equation}\label{qubit1}
f^{qb}_{\rm flux}(\phi,x)=\pm2 \pi \left(M I_p/\Phi_0\right) \delta'(x) \, ,
\end{equation}
where the different signs, $\pm$, correspond to the two persistent current states $|0\rangle$ and $|1\rangle$ of the qubit, respectively.

For a qubit remaining at the symmetry point, $\epsilon_0=0$, $g\ll\Delta$,  we obtain an inductive type interaction, and the perturbation is given by
\begin{equation}\label{qubit2}
f^{qb}_{\rm ind}(\phi,x)= \pm\frac{4 I_p^2 M^2/\Delta}{ L \lambda_J/a} \frac{\partial}{\partial x}\left[ \delta(x) \phi_x
\right].
\end{equation}
Herethe different signs, $\pm$, correspond to the energy
eigenstates of the qubit at the symmetry point,
$|\pm\rangle=(1/\sqrt{2})(|0\rangle\pm|1\rangle)$.

In the limit  $\alpha=j_e=f^{qb}=0$ the exact soliton solution of Eq.~(\ref{sine-Gordon perturbed}) has the form
\begin{equation}\label{soliton}
\phi_0(x,t)=4 \tan^{-1} \left[\exp\left(±\frac{x- u t- x_0}{(1-u^2)^{1/2}}\right)\right],
\end{equation}
where the positive (negative) sign corresponds to a fluxon (antifluxon), $x_0$ is the initial position of the soliton, and $u$ its velocity, which can take any value between $-1$ and $1$.
To analyze the fluxons dynamics in the JTL following from
Eq.~(\ref{sine-Gordon perturbed}) in the general case we make use of the collective coordinate perturbation theory developed by McLaughlin and
Scott~\cite{McLaughlin}. It is based on the assumption that
$j_e$, $\alpha \dot \phi$, $f^{qb}(\phi,x)\ll1$, which allows writing the fluxon solution in the form
\begin{equation}\label{soliton2}
\phi_0(x,t;u,X)=4 \tan^{-1} \left[\exp\left(\frac{x-X(t)}{(1-u^2(t))^{1/2}}\right)\right].
\end{equation}
For simplicity we consider here the case where only one fluxon is present in the system.

Without coupling to the qubit, $f^{qb}=0$, the stationary velocity of a fluxon
can be derived from the power balance equation~\cite{McLaughlin},
and is given by
\begin{equation}\label{t. velocity}
u_0=\left( 1+\left(\frac{4 \alpha}{\pi j_e}\right)^2\right)^{-1/2}.
\end{equation}
When the qubit is coupled to the JTL, for arbitrary form of the perturbation $f^{qb}(\phi,x)$, 
the variation parameters $X(t)$ and $u(t)$ obey the  differential
equations \cite{McLaughlin}
\begin{eqnarray}
\frac{du}{dt}&=&\frac{1}{4}\pi j_e (1-u^2)^{3/2}-\alpha u (1-u^2)-\frac{1}{4}(1-u^2)\nonumber\\
& &\times\int_{-\infty}^{\infty}f^{qb}\left(\phi_0(\Theta)\right) {\rm sech} \Theta \; dx,\label{u,X equts1}\\
\frac{dX}{dt}&=&u-\frac{1}{4}u (1-u^2)^{1/2}\nonumber\\
& &\times\int_{-\infty}^{\infty}f^{qb}\left( \phi_0(\Theta)\right) \Theta \; {\rm sech} \Theta \; dx,
\label{u,X equts2}
\end{eqnarray}
where $\Theta=(x-X(t))/(1-u^2)^{1/2}$. In the following section
Eqs.~(\ref{u,X equts1}) and (\ref{u,X equts2}) will be solved for both forms of the qubit perturbation $f^{qb}$ given in Eqs.~(\ref{qubit1}) or (\ref{qubit2}).

\begin{table*}
\caption{\label{parameters}Parameters of the JTL and the persistent current qubit.}
\begin{ruledtabular}
\begin{tabular}{ccccccccccccc}
 \multicolumn{5}{c}{Parameters of JTL~\cite{Averin, Savin}}&$\mid$&\multicolumn{2}{c}{Qubit-JTL coupling, k} &$\mid$&\multicolumn{4}{c}{Parameters of persistent current qubit~\cite{qubit}}\\\hline
$I_c$($\mu$A) &$J_c$(A/cm$^2$) &  $\lambda_J$($a$) & $L$(pH) & $\omega_p/2\pi$(GHz) &$\mid$&$g\gg\Delta$& $g\ll\Delta$ &$\mid$&$\Delta/2\pi$(GHz) &$\epsilon_{max}/2\pi$(GHz)&$L_{qb}$(pH)&$I_p$(nA)\\
0.6          &30               &  2                 & 137     & 24  &$\mid$& 0.96 & 0.05 &$\mid$&5.5          & 55                & 5           & 300\\
2            &30               & 2                  & 41      & 24  &$\mid$& 0.53 & 0.025&$\mid$&&&&
\end{tabular}
\end{ruledtabular}
\end{table*}

\section{Delay times.\label{delay time section}}
\subsection{Qubit far from the Symmetry Point}
We first consider the situation where the qubit is prepared far from the symmetry point, $\epsilon_0\gg\Delta$. In this case the eigenstates of the qubit are the persistent current states, $|0\rangle$ and $|1\rangle$. For each of them a magnetic flux $\pm M I_p$ penetrates the cell of the JTL which is inductively coupled to the qubit. If the size of the qubit 
 is much less than the Josephson penetration depth $\lambda_J$
the external magnetic flux in the JTL can be written as
\begin{equation}
\Phi^e(x)=\pm MI_p \theta(x),
\end{equation}
where $\theta(x)$ is the step function, and we assumed the qubit to be located at $x=0$.
The corresponding perturbation term in the equation of motion (\ref{sine-Gordon perturbed}) can be written  as
\begin{equation}\label{f^qb_1}
f^{qb}_{\rm flux}(\phi,x)=\pm\phi^q_1 \delta'(x),
\end{equation}
where $\phi^q_1= 2 \pi MI_p/\Phi_0$ is the dimensionless qubit-JTL
coupling. A thorough experimental and theoretical study of the fluxon
dynamics described by the sine-Gordon equation with delta-function terms
was performed in Refs.~\cite{Ustinov, Malomed} for long annular
Josephson junctions. Propagation of the charge soliton in 1D array of serially coupled Josephson junction was studied in Ref.~\cite{charge soliton}.

The dimensionless coupling
$\phi^q_1$ depends linearly on the mutual inductance $M$, which for typical experimental  conditions is much lower than $1$. Increasing the coupling coefficient  $k$  between the qubit loop and the JTL cell enhances the influence of the qubit on the fluxon dynamics in the JTL, but at the same time increases the back-action of the fluxon on the qubit.
In particular, the fluxon induces a magnetic flux in the qubit loop, which shifts the working point of the qubit. The persistent current qubit is only well defined when the external flux in the qubit loop is close to the symmetry point $\Phi_0/2$.
For larger values of the external magnetic flux the form of qubit-JTL perturbation (\ref{f^qb_1}) is not valid and a more elaborated model should be used. Moreover, large deviations of the working point can lead to non-adiabatic transitions to higher levels, which also makes the two-level model invalid. In the following we require $MI_p I_L|_{max} < \epsilon_{max}/2$, where
$\epsilon_{max}$ is the maximum value of the energy bias between two
persistent current states for which our system is still regarded as the persistent current qubit.
This condition translates to the following limitation for the dimensionless coupling 
\begin{equation}\label{condition}
  \phi_1^q < \frac{\epsilon_{max}}{\Phi_0^2 a/(\pi^2 L\lambda_J)},
\end{equation}
and for the coupling coefficient $k=M/\sqrt{L L_{qb}}$,
\begin{equation}\label{condition k}
  k<\frac{\pi}{2}\left(\frac{\epsilon_{max}\lambda_J}{I_p \Phi_0 a}\right)\left(\frac{L}{L_{qb}}\right)^{1/2}.
\end{equation}
The right hand side of Eq. (\ref{condition}) is the ratio between the qubit energy splitting and the magnetic energy of a fluxon $\Phi_0^2 a/(\pi^2 L \lambda_J)$.
It demonstrates that the efficiency of the measurement is low if the magnetic energy of a soliton is much higher than the qubit energy, because one needs to decrease the coupling coefficient $k$ according to (\ref{condition k}).

The magnetic energy of a soliton in the JTL can be reduced by
decreasing the critical current $I_c$ of the junctions.
We evaluate the delay times for two specific values of the critical current,
$I_c=0.6~\mu$A and $I_c=2~\mu$A. Given these values the cell inductance $L$ has to be chosen appropriate to yield suitable values
of the Josephson penetration depth $\lambda_J$. This length should be large  to minimize effects of the discreteness of the JTL, $\lambda_J\geq2a$. On the other hand, it is useful to have $\lambda_J$ as small as possible to decrease the magnetic energy of a fluxon and to keep the number of Josephson junctions in the JTL within practical limits. Based on these considerations~\cite{Averin} we choose $\lambda_J=2a$.  The parameters of the JTL and the qubit which were used in the calculations are listed in  Table~\ref{parameters}.


From (\ref{u,X equts1}-\ref{f^qb_1}) we obtain
\begin{eqnarray}
\frac{du}{dt}&=&\frac{1}{4}\pi j_e (1-u^2)^{3/2}-\alpha u (1-u^2)\nonumber\\
\label{F_uf1}
& &\pm\frac{1}{4}(1-u^2)^{1/2} \phi^q_1 {\rm sech}^2 \Theta_0 {\rm sh}\Theta_0, \\
\label{F_uf2}
\frac{dX}{dt}&=&u\pm\frac{1}{4} u\phi^q_1 \left( {\rm cosh} \Theta_0
-\Theta_0 {\rm sh} \Theta_0 \right) {\rm sech}^2\Theta_0.
\end{eqnarray}
where $\Theta_0=X/(1-u^2)^{1/2}$.

In the  ballistic regime, $j_e=0$ and $\alpha=0$, we can integrate (\ref{F_uf1}, \ref{F_uf2}) analytically and obtain the fluxon velocity $u$ and  center coordinate $X$ as functions of  the parameter $\Theta_0$ for the qubit prepared in the eigenstates $|0\rangle$:
\begin{eqnarray}\label{u1}
u(\Theta_0)&=&\left[ 1-(1-u_0^2)\left( \frac{\phi^q_1+4{\rm cosh}\Theta_0}{4{\rm cosh}\Theta_0}\right)^2 \right]^{1/2},\\
X(\Theta_0)&=&\Theta_0 (1-u_0^2)^{1/2}\left(1+\frac{\phi^q_1}{4{\rm cosh}\Theta_0} \right) .
\end{eqnarray}
The solution for the state $|1\rangle$ is obtained by replacing $\phi^q_1$ by $-\phi^q_1$.

\begin{figure*}[t]
\includegraphics[width=18cm, trim=1cm 1cm 1cm 1cm]
{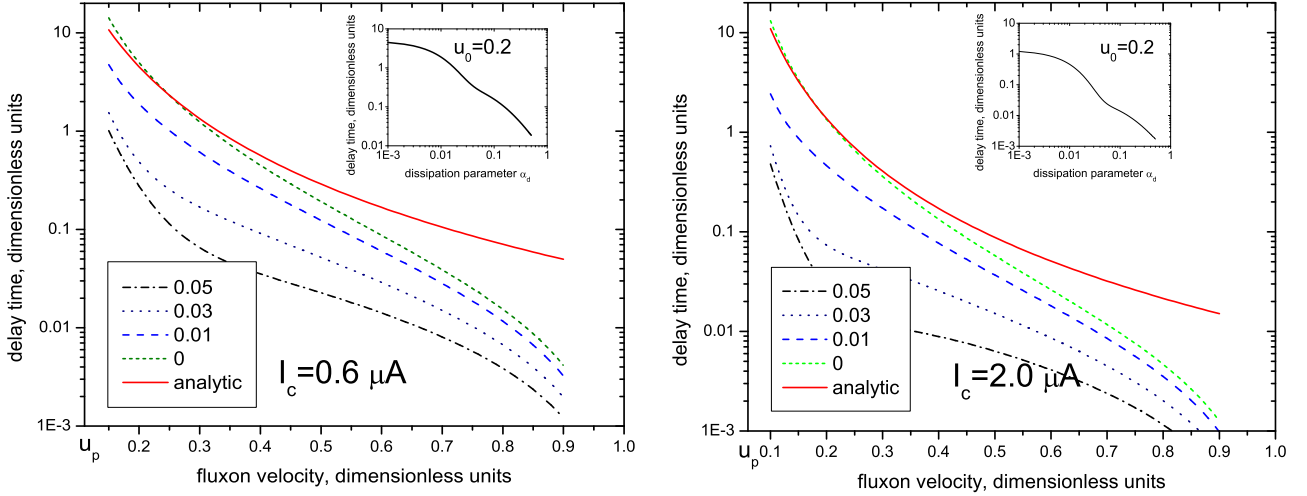} \caption{Delay time of a soliton induced by the qubit
as a function of initial soliton velocity $u_0$ when the qubit is far
from the symmetry point, $\epsilon_0\gg\Delta$. The velocity and time are
measured in units of $\lambda_J \omega_p$ and $\omega_p^{-1}$,
respectively. Four curves are obtained numerically from
Eqs.~(\ref{u,X equts1}, \ref{u,X equts2}) for different
values of the dissipation strength $\alpha$ (shown in the legend
box). The red solid line shows the analytical solution
(\ref{delay1}). The parameters of the JTL and the persistent current
qubit are listed in Table~\ref{parameters}. The inset shows the
dependence of the delay time on the dissipation strength
$\alpha$ for the initial soliton velocity $u_0=0.2$.} \label{delay
time flux}
\end{figure*}

To proceed we first consider the situation, where the soliton is able to pass the potential and, therefore, we can always introduce the delay time caused by the perturbation due to the qubit as
\begin{eqnarray}
t_d &=& |t_{|0\rangle}-t_{|1\rangle}|= 2|t_{|0\rangle}-t_0|\nonumber\\
\label{delay time}
& &=\frac{2}{u_0}\int_{-\infty}^{\infty}\left|\frac{u_0-u(X)}{u(X) } \right| dX.
\end{eqnarray}
Here $t_{|0\rangle}$ and $t_{|1\rangle}$ are the propagation times corresponding to the two persistent current states of the qubit, and $t_0$ is the value without qubit interaction. 
The factor $2$ in Eq.~(\ref{delay time}) reflects the fact that the perturbation $f^{qb}_{\rm flux}$ has the same magnitude but different signs for the two  persistent current  states.
When the velocity is only slightly perturbed by the qubit, $\varepsilon_1=(1-u_0^2)\phi^q_1/(4 u_0^2)\ll1$, we find
\begin{equation}
u(\Theta_0)=u_0 \left( 1-\varepsilon_1{\rm sech}\Theta_0\right)+o(\varepsilon_1^2).
\end{equation}
The delay time (\ref{delay time}) can then be evaluated as
\begin{equation}\label{delay1}
t_{d}=\frac{2\varepsilon_1}{u_0^2}\int_{-\infty}^{\infty} \varepsilon_1{\rm sech}\left( \frac{X}{(1-u_0^2)^{1/2}}\right)  dX
\simeq\frac{\pi \phi^q_1}{2 u_0^3},
\end{equation}
where the last part of the equation describes the ``non-relativistic case" $u_0\ll1$.

From (\ref{u1}) we see that for $\phi^q_1/4>(1-u_0^2)^{-1/2}-1$  the soliton does not have enough kinetic energy to pass
the potential barrier induced by the the qubit in the state $|0\rangle$. After approaching the qubit the soliton
will be reflected by the barrier.
Due to the combination of dissipation and driving the soliton
will oscillate and eventually stop at the ``pinning point" where $u=0$ and $\dot X=0$.
The pinning of a fluxon and can also be used for the transmission detection
mode of measurement, but will not be analyzed here further.

The results of a numerical evaluation of the delay times for the JTL with dissipation
are presented in Fig.~\ref{delay time flux}.  As expected the analytical formula
(\ref{delay1}) gives a good estimate even for intermediate values
of the soliton velocity $u_p\ll u_0\ll 1 $, where $u_p$ is the
minimal possible velocity for which a fluxon is still able to pass the potential barrier created by the qubit.

\begin{figure}[t]
\includegraphics[width=8cm, trim=1cm 1cm 1cm 1cm]
{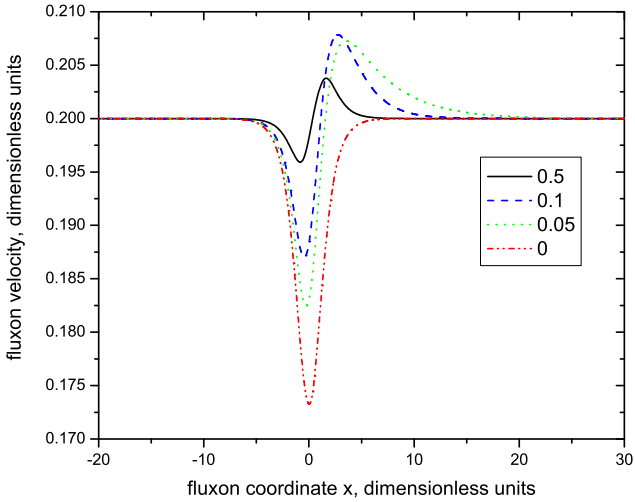} \caption{Fluxon velocity as function of its coordinate
for different
values of the dissipation strength $\alpha$ (shown in the legend
box)}. \label{u-x}
\end{figure}

Another observation is, that  for fixed value of the soliton
velocity $u_0$ the delay time decreases with increasing strength of the dissipation  $\alpha$. This property is displayed by Fig.~\ref{u-x}, which shows the velocity as a function of the soliton coordinate
for different values of $\alpha$. One can see that with  increasing  dissipation the fluxon velocity
 deviates from the dissipationless solution (\ref{u,X equts1},\ref{u,X equts2}). The initial decline of the fluxon velocity becomes
 less effective and is also partially compensated by the following acceleration of the fluxon. The compensation is the more complete
 the higher is the dissipation, which leads to zero delay times in the limit of strong dissipation.
This behavior is typical for the particle driven in viscous
media and can be understood from power balance considerations.
Let us consider the case when  the qubit induces a positive potential
barrier for a fluxon. When a fluxon is moving with stationary
velocity far from the qubit its energy gain and losses are equal.
Approaching the potential barrier the velocity of a fluxon 
decreases, which leads to a reduction in energy losses due to
dissipation. However, the gain of energy due to driving stays on the
same level. As a result a fluxon receives an excess amount of energy, which later
leads to an increase of the fluxon velocity above the initial value. Thus, we can use the
dissipationless solution as an approximation for evaluating the
delay times of a fluxon only if $\alpha<\phi^q_1$. For
$\alpha>\phi^q_1$ the delay times degrade rapidly with increasing
$\alpha$.  We expect that the delay times can be reliably
detected only if $\alpha\lesssim \phi^q_1$. This establishes
another limitation to the parameters of the JTL for the case
of very weak qubit-JTL coupling.

Next we consider the situation when the qubit is initially prepared at the symmetry point but shifted far from it by the moving fluxon. In what follows we will show that this regime is approximately equivalent to that when the qubit is kept far from the symmetry point all the time, and we can use the previous results for the delay time.
Indeed for typical parameters of the JTL and chosen inductive coupling we have $g\gg \Delta$, which proves that even if the qubit
is initially prepared at the symmetry point, $\epsilon_0=0$, it is shifted far
from it by the flux induced in the qubit loop by the moving fluxon. At the symmetry point the persistent current in the qubit loop is zero, and the qubit induces no magnetic flux in the JTL cell, $I_L(t)=0$. As soon as the approaching fluxon shifts the qubit from the symmetry point, $2MI_p I_L(t)\sim \Delta$, the persistent current is no longer zero. The maximum value of the persistent current, $I_p$, is reached once the qubit is pushed far away from the symmetry point, $2I_pMI_L(t)\gg \Delta$.  When the fluxon moves away the qubit returns to the symmetry point with zero persistent current in its loop. These considerations, of course, are valid only if the qubit was prepared in an energy eigenstate and if the shift by the fluxon occurs adiabatically. The qubit affects the fluxon dynamics only during the period of time $\sim\lambda_J/u_0$ when the fluxon is moving close to the qubit.  Most of this time the  qubit is already shifted far from  the symmetry point and induces the perturbation to the fluxon motion according to Eq.~(\ref{qubit1}). The quick switching of the persistent current from zero to $I_p$ and back in the initial and final stages of the qubit-JTL magnetic flux exchange yield only a small contribution to the overall delay time experienced by the fluxon. Therefore,  we can use the calculated delay times shown in Fig.~\ref{delay time flux} even if the qubit is initially prepared at the symmetry point for $g\gg \Delta$. However, we need to evaluate the probability of  non-adiabatic transitions of the qubit to another eigenstate, which creates an additional source of error for the measurement. These results will be presented in Section~\ref{backaction}.

\subsection{Qubit at the Symmetry Point}
\begin{figure*}
\includegraphics[width=18cm, trim=1cm 1cm 1cm 1cm]{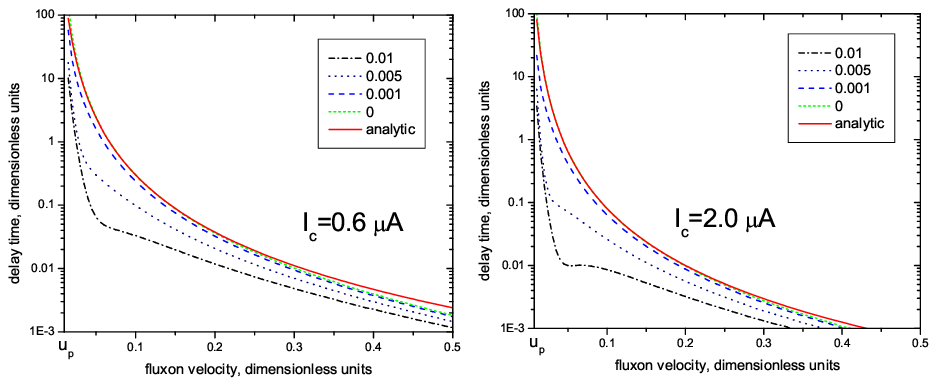} \caption{Delay time of a soliton induced by the qubit as
function of the initial soliton velocity $u_0$ when the qubit remains at the
symmetry point $\epsilon_0=0$. The velocity and time are measured in units of $\lambda_J \omega_p$ and $\omega_p^{-1}$, respectively.
Four curves are numerical solutions of Eqs.~(\ref{u,X equts1}, \ref{u,X
equts2}) for different values of the dissipation strength
$\alpha$ (shown in the legend box). The red solid curve shows the
analytical solution (\ref{delay time2}). The parameters of the JTL and persistent current qubit are given in Table~\ref{parameters}. The coupling coefficient between the qubit loop and the cell of the JTL is $k=0.05$ which leads to the dimensionless coupling  $\phi^q_1=3\times10^{-4}$ in the sine-Gordon equation. The value of the coupling coefficient is reduced in order to keep the qubit all the time at the symmetry working point. } \label{delay time inductive}
\end{figure*}

In this subsection we consider the situation where the qubit is prepared initially at the symmetry point $\epsilon_0=0$ and remains nearby during the measurement. This requires that the coupling between qubit and JTL is
weak and $g\ll\Delta$ to ensure that a soliton does not shift the qubit
from the symmetry point significantly.
As a consequence the expectation value of the flux in the energy
eigenstates of the qubit is  close to zero. If the coupling
term~(\ref{H_I})
varies slowly  we can treat the sum $H_{qb}+H_I$ in an adiabatic
approximation and diagonalize  it to obtain
\begin{eqnarray}\label{inductive H2}
H_{qb}+H_I&=&\rho_z\sqrt{\Delta^2/4+M^2I_p^2I_L^2(t)}\\
\nonumber
& &\simeq\frac{\Delta}{2}\rho_z+\frac{L_{eff}I_L^2(t)}{2}\rho_z=H_{qb}^{adiab}+H_I^{adiab},
\end{eqnarray}
where $L_{eff}=2 I_p^2 M^2/\Delta$.
The approximation is valid if $P\ll1$, where $P$ is the probability of  Landau-Zener
transition between energy eigenstates due to time dependence of $I_L(t)$, which is calculated in the next section.

The interaction term $H_I^{adiab}$ indicates that the qubit-JTL
interaction in the adiabatic approximation is equivalent to the change of the inductance of
the JTL cell which is coupled to the qubit by the additional value of $±L_{eff}$ whose sign depends
on the energy eigenstate of the qubit. This property can be used
for the read out of the qubit when it is always kept at the
symmetry point $\Phi=\Phi_0/2$.

The qubit-JTL interaction $H_I^{adiab}$ leads to the following perturbation term in the sine-Gordon equation (\ref{sine-Gordon perturbed})
\begin{equation}\label{perturbation induct}
f^{qb}_{\rm ind}(\phi,x)=\pm\phi^q_2 \frac{\partial}{\partial x}\left[ \delta(x) \phi_x \right],
\end{equation}
where $\phi^q_2=L^{eff}a /(L \lambda_J)$ is the corresponding dimensionless qubit-JTL coupling and $\pm$ correspond to the eigenstates $|\pm\rangle=(1/\sqrt{2})(|0\rangle\pm|1\rangle)$.

The Hamiltonian (\ref{inductive H2}) is valid only for $g\ll\Delta$,
which establishes the condition for the maximum possible
dimensionless coupling
\begin{equation}\label{condition3}
  \phi_2^q\ll \frac{\Delta/2}{\Phi_0^2 a/(\pi^2 L\lambda_j)},
\end{equation}
and coupling coefficient
\begin{equation}\label{condition3-1}
  k\ll\frac{\pi}{2}\left(\frac{\Delta\lambda_J}{I_p \Phi_0 a}\right)\left(\frac{L}{L_{qb}}\right)^{1/2}.
\end{equation}

In order to keep the qubit at the symmetry point for the chosen parameters of the qubit and the JTL (shown in Table~\ref{parameters})  the coupling coefficient should satisfy: $k\ll0.096$ for $I_c=0.6~\mu$A and $k\ll0.053$  for $I_c=2.0~\mu$A. For the calculation we take $k=0.05$ and $k=0.025$ which leads to $\phi^q_2=3\times10^{-4}$ and $\phi^q_2=8\times10^{-5}$, respectively.

From (\ref{u,X equts1},\ref{u,X equts2}) and (\ref{perturbation induct}) we obtain
\begin{eqnarray}
\frac{du}{dt}&=&\frac{1}{4}\pi j_e (1-u^2)^{3/2}-\alpha u (1-u^2)\nonumber\\
\label{F_u induc}
& &\pm\frac{\phi^q_2}{2} {\rm sech}^2 \Theta_0{\rm tanh}\Theta_0,\\
\frac{dX}{dt}&=&u\pm\frac{\phi^q_2}{2} \frac{u \;{\rm sech}^2\Theta_0}{(1-u^2)^{1/2}}   \left( 1-\Theta_0  {\rm tanh}\Theta_0\right).
\end{eqnarray}
For $\alpha=j_e=0$ in the non-relativistic limit $u\ll1$ one can evaluate for the qubit in the eigenstate $|+\rangle$
\begin{equation}
u^2=u_0^2-\frac{\phi^q_2}{2}{\rm sech}^2 X,
\end{equation}
and
\begin{equation}
u\simeq u_0\left(1-\varepsilon_2{\rm sech}^2 X \right)+o(\varepsilon_2^2),
\end{equation}
where $\varepsilon_2=\phi^q_2/(4 u_0^2)$. The pinning of the fluxon occurs if $u_0^2<\phi^q_2/2$.
We can calculate the delay time according to (\ref{delay time}) as
\begin{equation}\label{delay time2}
t_{d}=\frac{2\varepsilon_2}{u_0}\int_{-\infty}^{\infty}{\rm sech}^2 X  dX=\frac{\phi^q_2}{u_0^3}.
\end{equation}

The results of the numerical evaluation of the delay times in this
regime are shown in Fig.~\ref{delay time inductive}.
For the same values of the fluxon speed the delay times are smaller as compared to Fig.~\ref{delay time flux}.
However, since the pinning of
a soliton occurs at slower velocities we can, in principle,
achieve larger values of the delay times.

\section{Back-action of a Fluxon on the Qubit}\label{backaction}

The back-action of a fluxon on the qubit produces several effects
including  dephasing of the qubit in the measurement
basis~\cite{Averin2} and a shift of the working point of the qubit.
Dephasing, i.e., the decay of the off-diagonal elements of the
density matrix of the qubit in the measurement basis,
does not affect the measurement outcomes studied here~\cite{dephasing} and is not considered  further.

In contrast, the shift
of the working point of the qubit can potentially ``destroy"  the
qubit and, furthermore, induce non-adiabatic transitions between the
qubit states. Both processes create errors in the measurement and should
be taken into account. By restricting the qubit-JTL
coupling according to (\ref{condition k}) or (\ref{condition3-1}) we
insure that the system remains a persistent
current qubit during the whole time of the measurement. In this section
we derive the probability of non-adiabatic transition between
eigenstates of the qubit due to a passing fluxon. It should be
noted that non-adiabatic transitions are a potential problem 
only if the qubit is
initially prepared at the symmetry point. Far from the symmetry point
the Hamiltonian of the qubit (\ref{total H1}) and the interaction
Hamiltonian (\ref{H_I}) approximately commute and  non-adiabatic
transitions are negligible.

In order to study  nonadiabatic transitions we assume that the qubit is initially prepared in the excited state $|-\rangle$ at the symmetry point $\epsilon_0=0$ where its unperturbed dynamics is described by the Hamiltonian $H_{qb}=-(\Delta/2)\sigma_x$.
For a fluxon passing the qubit the time-dependent qubit-JTL coupling can be written according to (\ref{H_I}) and (\ref{soliton}) as
\begin{eqnarray}
\label{H_I(t)}
H_I&=&MI_pI_L(t)\sigma_z\\
\nonumber&=&-\frac{M I_p\Phi_0}{\pi L (\lambda_J/
a)\sqrt{1-u^2_0}}\; 
{\rm sech} \left(\frac{u_0 t}{\sqrt{1-u_0^2}}\right)\sigma_z.
\end{eqnarray}
Here we neglected the action of the qubit on the fluxon.
The transition probability from the excited state $|-\rangle$ to the ground state $|+\rangle$ induced by one passing fluxon is then given by~\cite{Zener}
\begin{eqnarray}
P=MI_p(2\hbar)^{-2}\left|
\frac{\sin A}{A}\int^{\infty}_{-\infty} I_L(t) e^{i t \Delta  /\hbar} dt\right|^2,\label{Prob}
\end{eqnarray}
where $A=MI_p(2\hbar)^{-2}\int^{\infty}_{-\infty} I_L(t) dt$.
After inserting (\ref{H_I(t)})  we obtain
\begin{equation}\label{Prob2}
P= \sin^2\left(\frac{I_p M \Phi_0}{L (\lambda_J/a)u_0 \hbar \omega_p}\right){\rm sech}^2  \left( \frac{\pi \sqrt{1-u_0^2}\Delta}{2
u_0 \omega_p}\right).
\end{equation}
We will use this result to evaluate the
measurement efficiency in  Section~\ref{discussions}.


\section{Soliton Jitter}

In order to be detected the delay time should exceed the time fluctuations induced by thermal noise in the JTL.
We can account for thermal noise by adding a stochastic term $\xi(x,t)$ to the sine-Gordon equation (\ref{sine-Gordon perturbed}),
\begin{equation}\label{sine-Gordon4}
\ddot \phi -\phi_{xx} +\sin \phi=j_e-\alpha \phi_t-\xi(x,t).
\end{equation}
The  noise is assumed to be white with autocorrelation function
\begin{equation}\label{n(x,t)}
\langle \xi(x,t)\xi(x',t')\rangle=16 \alpha(k_B T/E_0)\delta(x-x')\delta(t-t').
\end{equation}
Here $E_0=8\left( \hbar J_c \lambda_J/2e\right)$ is the rest energy
of a soliton, and the coefficient $16 \alpha(k_B T/E_0)$ is
fixed by the fluctuation-dissipation theorem~\cite{Salerno}.  The noise spectral density for the fluxon velocity is given by~\cite{joergensen,Salerno}
\begin{equation}\label{S_u}
S_{\Delta u}(\omega)=\frac{2 \alpha k_B T}{E_0}\frac{(1-u_0^2)^{5/2}}{\alpha^2+\omega^2},
\end{equation}
where $\Delta u(t)=u(t)-u_0$. Starting from the initial condition that at $t=0$ the velocity of the fluxon is  controlled and equal to $u_0$  we find the velocity autocorrelation function to be given by
\begin{eqnarray}
\langle \Delta u(t_1) \Delta u(t_2) \rangle&=&\frac{k_B T}{E_0}(1-u_0^2)^{5/2} \nonumber\\
& &\times\left(e^{-\alpha |t_1-t_2|}-e^{-\alpha |t_1+t_2|}\right).\label{uu}
\end{eqnarray}
The
uncertainty in the velocity of the fluxon leads to an uncertainty
of its coordinate $\Delta X(t)=\int_0^t \Delta u(t')dt'$ and propagation time  $\Delta t=\Delta X/u_0$. The time jitter of
the fluxon, defined as the standard deviation of the propagation time
$\delta t\equiv\left( \langle \Delta t^2\rangle\right)^{1/2}$, is
then given by
\begin{eqnarray}
\delta t&=&\left( \frac{2 k_B T}{\alpha^2
E_0}\right)^{1/2}\frac{(1-u_0^2)^{5/4}}{u_0}\nonumber\\
& &\times\left(t\alpha+e^{-\alpha t}-1-\frac{1}{2}\left(1-e^{-\alpha t}\right)^2\right)^{1/2}.\label{jitter}
\end{eqnarray}
Limiting cases of Eq.~(\ref{jitter}) are $\delta t\propto
t^{1/2}$ in the diffusive regime $\alpha t\gg 1$, and  $\delta
t\propto t^{3/2}$ in the ballistic regime $\alpha t\ll 1$. The time
jitter (\ref{jitter}) leads to further errors in the measurement
procedure, which will be analyzed in section VII.

\begin{figure*}[t]
  \begin{center}
    \scalebox{1}{\includegraphics[width =5cm]{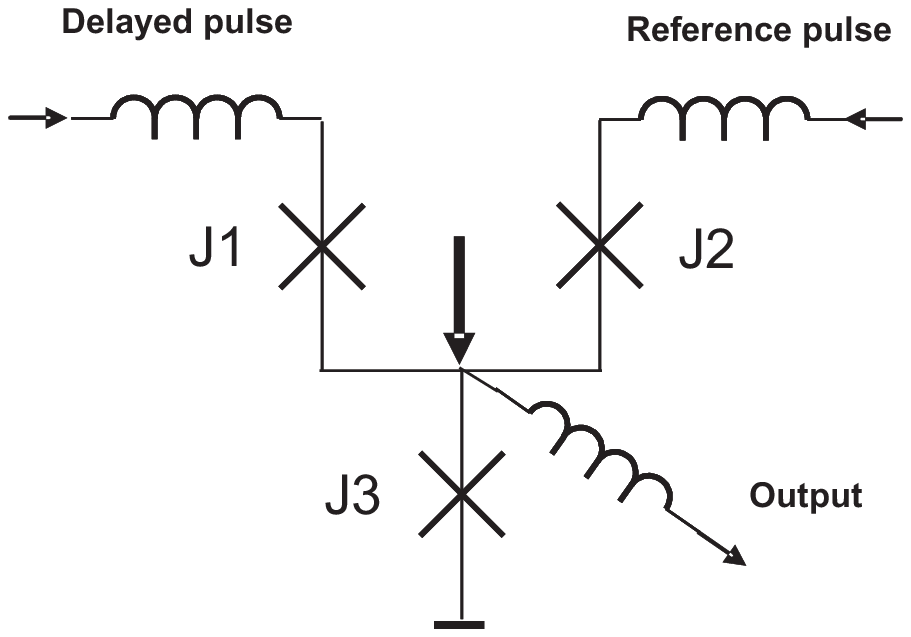}}(a)
    \scalebox{1}{\includegraphics[width =7cm, trim= 0cm 10cm 0cm 8cm]{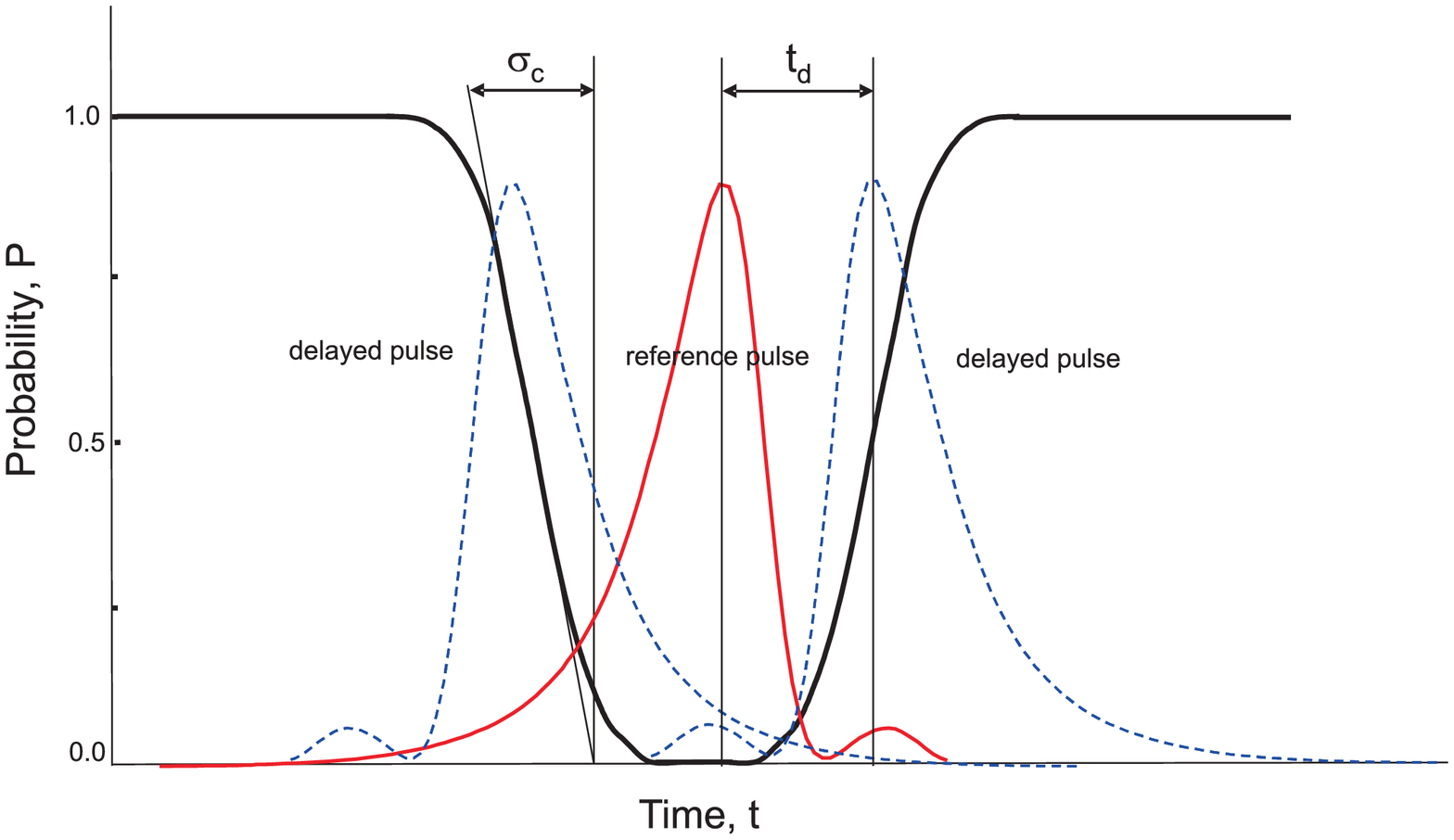}}  (b)
    \caption{(a) Schematic of RSFQ confluence buffer, and (b) the probability of double switching of output junction J3.}
    \label{merger}
  \end{center}
\end{figure*}

\section{RSFQ delay detector}

Before turning to a quantitative analysis of the measurement errors we describe in this section an RSFQ delay detector that is need to measures the time between two SFQ pulses propagating through a JTL. We first evaluate the time resolution of a single RSFQ decision gate and the time resolution of the improved detector based on a time vernier. Based on realistic parameters we estimate the hardware complexity of the
detector and discuss possible designs of the circuit presently developed experimentally~\cite{VTT}.

The operation of any RSFQ gate is based on the time resolution between SFQ pulses. The gates produce
binary output depending on the relative time between either clock and data pulses or between two
input data pulses. The simplest RSFQ gate that can be used as a decision circuit is the
asynchronous {\bf OR} or confluence buffer shown in Fig. \ref{merger}. The figure also illustrates the
operation of the confluence buffer as a time detector. A single SFQ pulse on either input produces
a SFQ pulse on output. Therefore, for delayed input pulses two output SFQ pulses are produced. In
case of simultaneous arrival the input pulses compensate each other resulting in a single SFQ pulse on
output. Such behavior of the RSFQ confluence buffer is well known and has been experimentally verified
many times, see for example Ref.~\cite{Bunyk97}.

In general, the time resolution of any RSFQ gate is determined by the  resolution of the balanced
comparator formed by two junctions connected in series. In the confluence buffer balanced
comparators are formed by junctions J1, J3 or J2, J3. The probability of junction switching in the
comparator obeys a normal distribution with $\sigma_c \approx 0.13 t_{SFQ}$, where $t_{SFQ}$ is the
width of SFQ pulse \cite{Ortlep06, Filippov99}. In time units normalized to $1/\omega_p$, the width of
the SFQ pulse is approximately

\begin{equation}
t_{SFQ} \approx 6\frac{1}{\sqrt{\beta_c}},
\end{equation}
where $\sqrt{\beta_c} = \omega_c / \omega_p$ is the McCumber-Stewart parameter. The corresponding
normalized one-sigma jitter of a single decision gate with $\beta_c=2$ is

\begin{equation}
\sigma_c = 0.13 t_{SFQ} \approx 0.78 \frac{1}{\sqrt{\beta_c}} \approx 0.55.
\end{equation}
For a measurement error of $10^{-2}$ the time resolution of a single RSFQ gate, $t_r^g$, is given by
$4\sigma_c$ and equal to $t_r^g = 2.2$. This time resolution does not dependent on the temperature
since the measurement involves symmetric stochastic processes of the switching of two balanced
junctions \cite{HerrIsec}.

\begin{figure}[t]
\centering
\includegraphics[width=70mm]{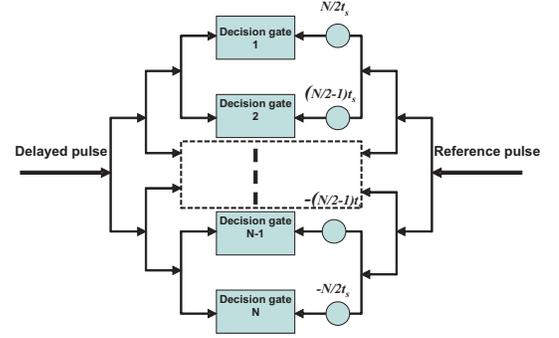}
\caption{Block diagram of a RSFQ time vernier.} \label{vernier}
\end{figure}

As it is shown in Section~\ref{delay time section}, the delay time between SFQ pulses that needs to be
detected is in the range of $t_d = 0.1 - 5$ depending on the  actual dissipation in long JTLs and the 
accuracy in fixing the initial fluxon velocity. In order to improve the time resolution of the RSFQ delay
detector time a vernier with $N$ decision gates can be used \cite{Kirichenko01,Fujimaki06}, the block
diagram of which is shown in Fig.~\ref{vernier}. After the JTLs, both pulses
propagate through the chain of splitters and arrive on a chain of $N$ decision gates with inputs
delayed in time by $t_d$. The operation principle of the time vernier is the same as that of a multibit
analog-to-digital converter where each bit compares the signal with a slightly shifted threshold. In
accordance with the standard theory for analog-to-digital converters the resolution improves as the square
root of the number of bits \cite{ADC}.

Time resolution of the vernier is $t_r^v =t_r^g/\sqrt{N}$ and relative time difference between
stages of the vernier is $t_s=t_r^g/N$.  The error of time measurements depends on the ratio
between time difference, $t_s$, and jitter accumulated in the pulse pathes, $\sigma_v$, 
\begin{equation}
P_{err} = \frac{1}{2}{\rm erfc}(\frac{t_s}{\sigma_v}).
\end{equation}

The accumulated jitter is 
\begin{equation}
\sigma_v = \sqrt{2*N_v}\sigma_J,
\end{equation}
where $N_v$ is number of Josephson junctions in each pulse path and $\sigma_J$ is a jitter per
Josephson junction. The factor two comes from the fact that there are two chains of splitters involved.
The number of Josephson junctions in the splitter chain grows like $N_v =N\log N$.   The single junction
jitter at 4.2 K is about $\sigma_J = 0.015 t_{SFQ}$ \cite{Ortlep06} and it scales as square root of
temperature and as square root of shunts resistors of the junction.

Typical operating bath temperatures for the qubit experiments are about 30\,mK. For process with operating frequency below 5\, GHz and assuming use of cooling fins
for thermalization of the hot electrons in the normal metal shunts,
the effective noise
temperature of the RSFQ gates can be reduced down to 80\,mK for a 30\,A/cm$^2$ process and to 30\,mK for a 
10\,A/cm$^2$ process \cite{Ohki06}. The temperature range
between 30-80\,mK corresponds to one-sigma jitter of $\sigma_J=0.0012 - 0.0004$ for a single
Josephson junction with $\beta_c = 2$.

For  measurement errors of $10^{-2}$ the time resolution of the RSFQ time vernier is given by
$4\sigma_v$ that results in the following relation for the optimum number of stages, 
\begin{equation}
\frac{4\sigma_c}{N} = 4\sqrt{2*N\log N}\sigma_J. \label{Eq.N}
\end{equation}
Equation (\ref{Eq.N}) gives $N = 40$ and corresponding $t_r^v= 0.33$ for 30\,mK (10 A/cm$^2$
process) and $N = 20$ and corresponding $t_r^v= 0.5$ for 80\,mK (30 A/cm$^2$ process).

\begin{figure}[hb]
\includegraphics[width=8cm, trim=1.5cm 0cm 0cm 2cm]{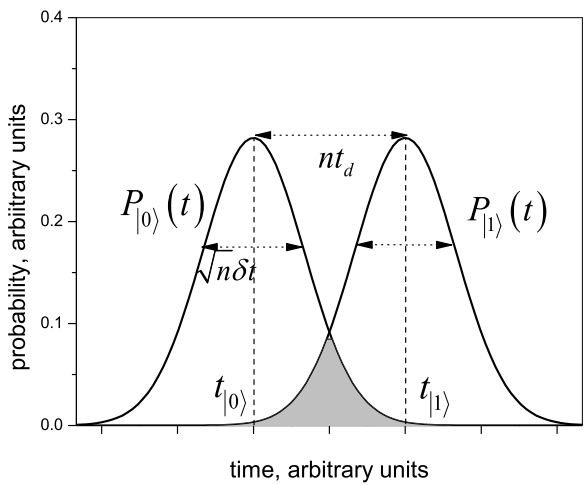}
\caption{Sketch of the distribution functions which describe the
propagation time of a soliton. The two functions $P_{|0\rangle}(t)$ and $P_{|1\rangle}(t)$ 
are peaked around the average values of the propagation times
$t_{|0\rangle,|1\rangle}$ corresponding to the different eigenstates of the qubit
$|0\rangle$ and $|1\rangle$. The standard deviation and peak separation are
given by $\sqrt{n}\, \delta t$ and $n\, t_d$, respectively. The overlap
of the distribution functions (grey color) indicates the regime
where the measurement yields an error.} \label{distribution}
\end{figure}

\begin{figure*}
\includegraphics[width=15cm]{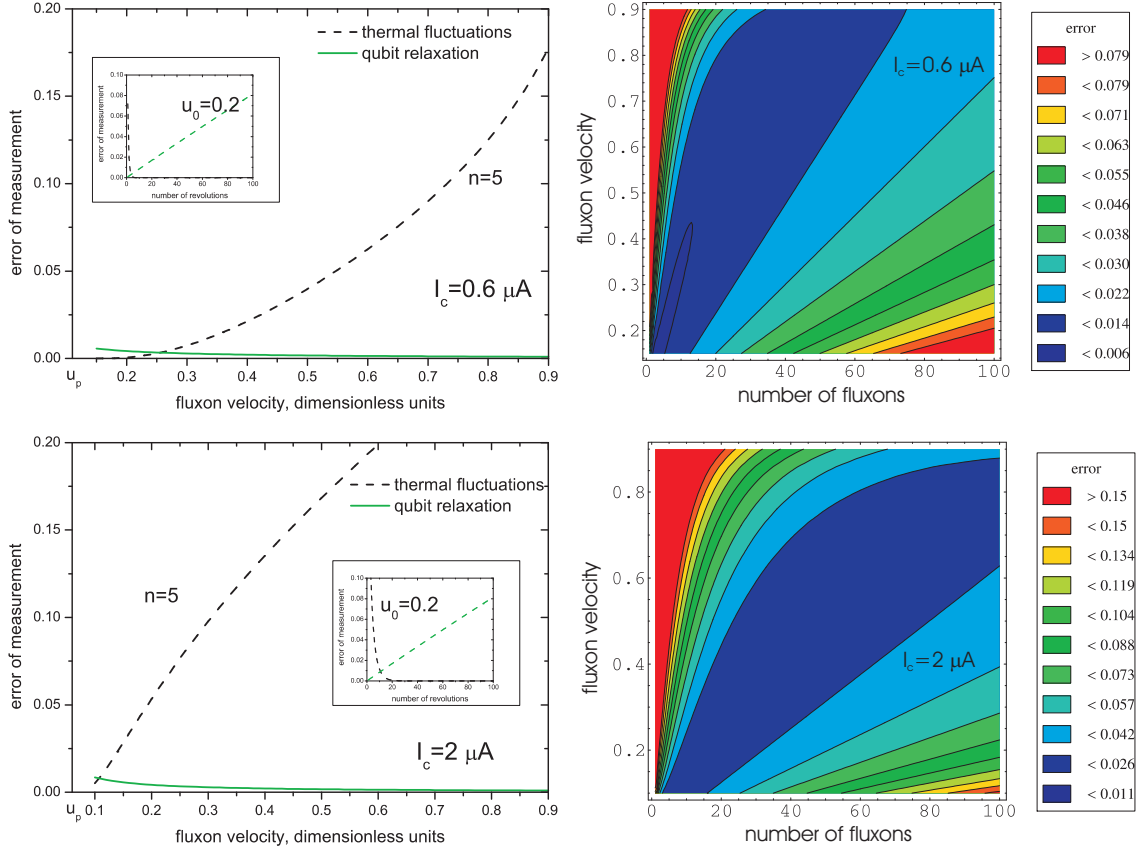}
\caption{Different contributions to the measurement error for
$L_r=25\lambda_J$, $T=20\;{\rm mK}$, $\alpha=0.001$. The upper
plots correspond to $I_c=0.6~\mu$A and $k=0.96$ while for the lower
plots $I_c=2~\mu$A and $k=0.53$. The qubit is initially prepared far
from the symmetry point. See text for detailed explanations.}
\label{error1}

\end{figure*}

\section{Measurement errors}\label{discussions}
In an experimental implementation the length $L_{JTL}$ of the JTL is
finite. For $L_{JTL}\gg \lambda_J$ the
delay time of a fluxon induced by the qubit is that of
the infinite JTL shown in Figs.~\ref{delay time flux} and \ref{delay
time inductive}. The time jitter of a fluxon at the end of the JTL
is given by Eq.~(\ref{jitter}) with $t=L_{JTL}/u_0$. Due to the jitter
the propagation time of a soliton will be scattered around the
average values $t_{|0\rangle}$ and $t_{|1\rangle}$ corresponding to the different energy eigenstates of the qubit $|0\rangle$ and $|1\rangle$ if the qubit is far from the symmetry point. (A similar analysis holds for distinguishing the states $|+\rangle$ and $|-\rangle$ if the qubit is at the symmetry
point.) We can introduce the corresponding distribution functions
$P_{|0\rangle}(t)$ and $P_{|1\rangle}(t)$ for the propagation time of a
soliton, which have maxima at $t_{|0\rangle}$ and $t_{|1\rangle}$, separated by the delay time $t_d$ and whose standard deviations are equal to the
jitter $\delta t$. The overlap of the two distribution functions
$P_{|0\rangle}(t)$ and $P_{|1\rangle}(t)$ characterizes the potential to distinguish the two eigenstates of the qubit. For
a normal distribution the area of
the overlap and the associated error of the measurement are
erfc$[t_d/\left(2\sqrt{2}\delta t\right)]$ and
(1/2)erfc$[t_d/\left(2\sqrt{2}\delta t\right)]$,
respectively. For experimentally relevant case of large separation of the distribution functions, $t_d\gg\delta t$, the error can be approximated by $\left(\sqrt{2} \delta t/t_d\right)\exp\left(-t_d^2/8\delta t^2\right)$. The ratio of the delay time to the jitter,  which
determines the error of measurement due to thermal fluctuations, can
be estimated for a JTL with low dissipation, $\alpha\ll\phi^q_{1,2}$, according to the expressions (\ref{delay1}), (\ref{delay time2}) and (\ref{jitter}) as
\begin{equation}\label{td/delta t}
\frac{t_{d}}{\delta t}\propto\frac{\pi \phi^q}{2
u_0^{3/2}}\left( \frac{E_0 \alpha}{2 k_B T L_{JTL}}\right)^{1/2}.
\end{equation}
This results suggests that in order to decrease the effect of thermal noise it is  favorable to use slower fluxons. On
the other hand, low velocity fluxons can be difficult to
produce and control technologically. To avoid pinning by the qubit we also cannot choose the speed of the fluxons too small.

For realistic experimental situations some of the magnetic flux
from the control circuit of the qubit $\Phi_{ext}$  penetrates the JTL
and create an additional potential barrier for fluxons. If
the external flux $\Phi_{ext}$ is much smaller than the flux
quantum $\Phi_0$ we can again use the collective perturbation theory
to analyze its effect on the delay time. According to the definition
(\ref{delay time}) there is no effect of $\Phi_{ext}$ on the difference in the delay times since the potential barrier does not depend on the qubit
quantum state. However, for $\Phi_{ext}> MI_p$ the pinning velocity
of a fluxon will be larger. This should be taken into account and
can affect the measurement because low fluxon velocities 
could be no longer accessible.

For further improvement of the signal-to-noise ratio one can use $n$
fluxons, sending them to the JTL one after another. After $n$ fluxon
passings the sum of the delay times  and separation of the peaks of
the distribution functions will scale linearly as $n\, t_d$, but the
 jitter will increase  only as $\sqrt{n}\,
\delta t$, and the resulting measurement error decreases
according to  $(1/2){\rm
erfc}[\sqrt{n}t_d/\left(2\sqrt{2}\delta t\right)]$.

\begin{figure*}
\includegraphics[width=15cm]{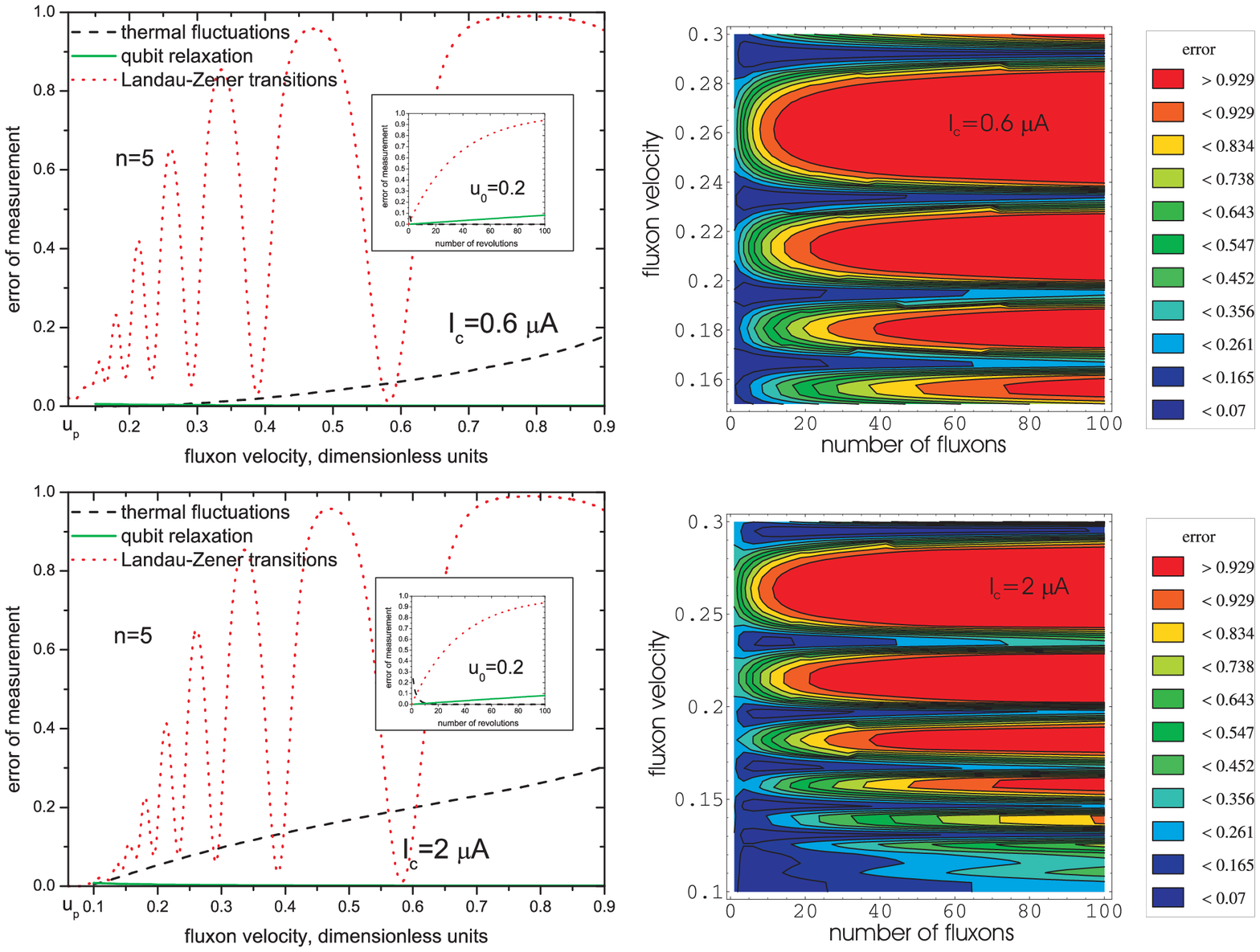}
\caption{Different contributions to the measurement error for
$L_{JTL}=25\lambda_J$, $T=20\;{\rm mK}$, $\alpha=0.001$. The upper
plots correspond to $I_c=0.6~\mu$A and $k=0.96$ while for the lower
plots $I_c=2~\mu$A and $k=0.53$. The qubit
is initially prepared at the symmetry point but is shifted far from it
by a moving soliton. See text for detailed explanations.}
\label{error2}
\end{figure*}

The qubit also experiences other sources of noise
during the measurement which are not related to the JTL. Their effect
 can be described phenomenologically  by
two characteristic time scales: the relaxation time $T_1$ and the
decoherence time $T_2$. The pure decoherence, i.e., the decay
of the off-diagonal elements of the density matrix of the qubit, has no
effect on the measurement outcome and is not relevant in our case.
The relaxation changes the probability distribution described by the
diagonal elements of the density matrix and, thus, affects the measurement.
For a qubit initially prepared in the
excited state, after $n$ passings fluxons the probability to
find the qubit in the ground state is
$1-\exp(-nL/(u_0 T_1))$. We take this probability as the approximate (upper bound) measure of the possible error of measurement due to relaxation of
the qubit.

\begin{table}[b]
\caption{\label{error 1 fluxon} Error of measurement by a single fluxon
for $L_{JTL}=25\lambda_J$, $T=20\;{\rm mK}$, $u_0=0.2$,
$\alpha=0.001$.}
\begin{ruledtabular}
\begin{tabular}{lccc}

& Variant   1 & Variant 2 & Variant 3\\
\hline
$I_c$($\mu$A)               & 0.6  & 2    & 0.6  \\
k                           & 0.93 & 0.53 & 0.05 \\
$t_d$($\omega_p^{-1}$)      & 4.5  & 1.2  & 0.04 \\
$\delta t$($\omega_p^{-1}$) & 1.5  & 0.84 & 1.5  \\
\hline
\multicolumn{4}{l}{Errors of measurement (in $\%$)}\\
\hline
jitter                      & 7    & 24   & 50   \\
relaxation                & 0.08 &0.08  & 0.08\\
LZ transitions, $\epsilon_0=0$& 2& 2& 6.8\\
\hline
total error,  $\epsilon_0\gg\Delta$& 7 &24 &-\\
total error, $\epsilon_0=0$ & 9 & 26 &55 \\
\end{tabular}
\end{ruledtabular}
\end{table}

Finally, if measure the qubit at the symmetry point
there is an additional error associated with 
Landau-Zener transitions between the qubit eigenstates.
Following considerations similar to the previous paragraph we define
the measure of the error due to Landau-Zener transitions as
$1-\exp(-n P)$ where probability $P$ is given by (\ref{Prob2}).

The efficiency of the qubit measurement by a single fluxon is given in
Table~\ref{error 1 fluxon}. The qubit can be measured at the symmetry point and far from it with
accuracy of approximately $90\%$ for $I_c=0.6~\mu$A and $75\%$ for $I_c=2~\mu$A.
To improve the accuracy one needs to use many fluxons.
Figs.~\ref{error1} and \ref{error2} show the error of the measurement
for the qubit initially prepared far from symmetry point and at the symmetry point, respectively. The
total errors of the measurements are shown on
the right contour plots as a function of the number of fluxons passings
 and the initial velocity of the fluxons. They are estimated as
$1-\exp\left[-{\rm erfc}(\sqrt{n}t_d/2\sqrt{2}\delta t)-nL/(u_0
T_1)\right]$ for Fig.~\ref{error1} and $1-\exp\left[-{\rm
erfc}(\sqrt{n}t_d/2\sqrt{2}\delta t)-nL/(u_0 T_1)-nP\right]$  for
Fig.~\ref{error2}.  The different contributions to the total error are
also plotted separately as a function of the fluxon velocity (the left
plots) and number of fluxon passings (the insets of the left plots).
The non-adiabatic transitions are suppressed if the qubit is prepared
far from the symmetry point, hence there is no curve
related to Landau-Zener transitions in Fig.~\ref{error1}. One can
see that for high fluxon velocities the measurement quality is
limited by the jitter and the Landau-Zener transition probability (if the qubit is at the symmetry point).
For low velocities the measurement time increases and the
relaxation of the qubit becomes also important. Under optimum
conditions the qubit can be measured with probability of error below $1\%$
far from symmetry point and below $7\%$ at the symmetry point
with the measurement time of approximately $4$~ns.

\begin{figure}
\includegraphics[width=8cm, trim=2cm 1cm 1cm 1cm]{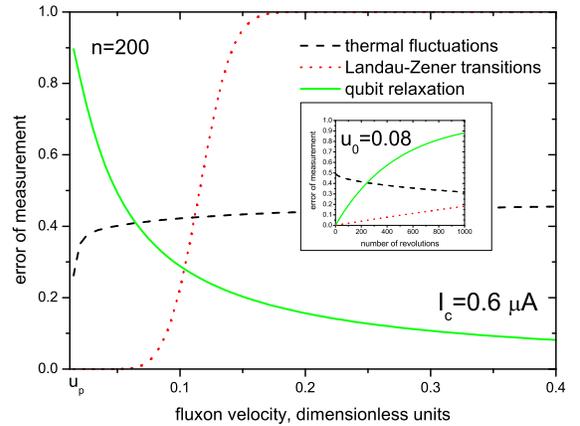}
\caption{Different contributions to the error of a measurement for
$L_{JTL}=25\lambda_J$, $T=20\;{\rm mK}$, $\alpha=0.001$,
$I_c=0.6~\mu$A and $k=0.05$. The qubit stays at the symmetry point
during the measurement.} \label{errorinduct}
\end{figure}

Fig.~\ref{errorinduct} shows the measurement error for
$L_{JTL}=25\lambda_J$, $T=20\;{\rm mK}$, $\alpha=0.001$ and
$k=0.05$. The qubit stays at the symmetry point during complete
measurement. The coupling coefficient is chosen small to prevent a shift of
the qubit from the symmetry point by the passing soliton. Thus, the
influence of the qubit on a soliton is also weak, which affects the
sensitivity of the measurement. From Fig.~\ref{errorinduct} it
becomes clear that in order to achieve better measurements we need
low soliton speed and a large number of fluxons passings. As a
result the time of measurement which is required to extract the
information about the qubit quantum state is comparable to the relaxation
time of the qubit. This makes it impossible to efficiently measure a
qubit in this regime for the chosen parameters of the JTL. A reduction of the critical current of the Josephson junctions may allow decreasing the
magnetic energy of the solitons and consequently increase the coupling
coefficient. In this case the accuracy of the measurement will be
improved while the qubit will remain all the time at the symmetry point.

\section{Summary}\label{summary}

In summary, we analyzed the measurement process of a quantum state
of a flux qubit by solitons propagating in an underdamped JTL. The coupling between the 
qubit and the JTL is inductive. We focused on the regime in which 
the information about the measured state is stored in the delay time 
of JTL solitons passing by (scattering from) the qubit. If the qubit 
is in an energy eigenstate far from the symmetry point its 
persistent current induces an external magnetic flux in the JTL 
which serves as a scattering potential for the JTL solitons. Two 
eigenstates induce different (opposite) scattering potentials, 
thus producing different delay times. A similar measurement mechanism works 
when the qubit is prepared at the symmetry point but, due to strong 
coupling, each passing soliton pushes the qubit  
far from the symmetry point.  

We have demonstrated that the delay times are longer for
lower magnetic energies and velocities of the fluxons and lesser
dissipation in the JTL. Since the Josephson penetration depth is
fixed by the design requirement $\lambda\approx2a$ the only
possibility to reduce the magnetic energy of the soliton is to
decrease the Josephson junction critical currents $I_c$. For low
critical currents $I_c=0.6~\mu$A and $I_c=2.0~\mu$A the delay
times are in the range $t_d=0.1-10\omega_p^{-1}$ which can be
reliably detected by the RSFQ delay detector. The major sources of
the measurement errors are the propagation time uncertainty of a
fluxon due to thermal noise in the JTL and the intrinsic qubit
relaxation. Non-adiabatic transitions between the energy
eigenstate induced by fluxons can cause an additional error if the
qubit is measured at the symmetry point. For a JTL consisting of
50 elementary cells, dissipation strength $\alpha=0.001$,
temperature $T=20$~mK, fluxon velocity $u_0=0.2$, and 
coupling coefficient $k\sim1$ the measurement errors of the
qubit by a single fluxon are $9\%$ and  $26\%$ for $I_c=0.6~\mu$A
and $I_c=2.0~\mu$A, respectively. In order to increase the
signal-to-noise ratio one can use many fluxons which results in
the improved accuracy of measurement exceeding $99\%$ for the qubit
far from the symmetry point and $90\%$  at the symmetry
point.

For weak qubit-JTL coupling, $k\ll0.01$, a qubit prepared
at the symmetry point stays there all the time and induces no
magnetic field in the JTL. In this case the measurement can be
based on the fluxon scattered by the potential associated with the
change of the effective inductance of that JTL cell which is
coupled to the qubit loop. We found that a further
reduction of the critical current of the Josephson junction is
required for an effective measurement in this regime.

\begin{acknowledgments}
We thank C. Hutter, A. Poenicke and M. Siegel for stimulating discussions and support.
The work was supported by the EU Specific Target Project RSFQubit. 
\end{acknowledgments}

\end{document}